# On the Hidden Transient Interphase in Metal Anodes: Dynamic Precipitation Controls Electrochemical Interfaces in Batteries

Stephen T. Fuller[1] and J.-X. Kent Zheng[1,2]*

*1. Department of Chemical Engineering, University of Texas at Austin, Austin, Texas 78712, USA*

*2. Texas Materials Institute, University of Texas at Austin, Austin, Texas 78712, USA*

**Abstract:** The Solid-Electrolyte Interphase (**SEI**) formed on a battery electrode has been a central area of research for decades. SEI consists of a variety of byproducts generated by spontaneous and electrochemical decomposition of the electrolyte. This thin, structurally complex layer profoundly impacts the electrochemical deposition morphology and stability of the metal in battery anodes. Departing from conventional approaches, we investigate metal dissolution—the reverse reaction of deposition—in battery environments using a state-of-the-art electroanalytical system combining a rotating-disk electrode and *in-operando* visualization. Our key finding is the presence of a Transient Solid-Electrolyte Interphase (**T-SEI**) that forms during fast discharging at high dissolution rates. We attribute T-SEI formation to transient local supersaturation and resultant electrolyte salt deposition. The T-SEI fundamentally alters the dissolution kinetics at the electrochemical interface, leading to a self-limiting morphological evolution and eventually yielding a flat, clean surface. Unlike a classical SEI formed due to electrolyte decomposition, the T-SEI is fully "relaxable" upon removal of the enforced dissolution current; That is, the T-SEI completely dissolves back into the electrolyte when rested. The formation of T-SEI, surprisingly, plays a critical role in the subsequent electrodeposition. When the metal is redeposited on a fully relaxed T-SEI surface, the morphology is remarkably different from that deposited on pristine or low-rate discharged metal electrodes. Electron backscatter diffraction analysis suggests the deposition occurs via growth of the original grains; this is in stark contrast to the isolated, new nuclei seen on standard metal electrodes without T-SEI formation. Using 3D profilometry, we observe a 42% reduction in surface roughness due to T-SEI formation. Our findings provide important insights into the electrochemical kinetics at the metal-electrolyte interface, particularly in concentrated or "water-in-salt" electrolytes that are close to the salt saturation limit. The results suggest a new dimension for electrochemical engineering in next-generation batteries cycled at high rates.

**Significance statement**: The interphase between the electrode and electrolyte critically influences battery performance. Traditionally, this solid interphase, formed by electrolyte decomposition, is considered persistent. Using a state-of-the-art rotating-disk electrode system with *operando* visualization, we uncover that rapid solute cation production can create local supersaturation, causing solute precipitation. This solute precipitation constitutes a novel transient Solid Electrolyte Interphase (T-SEI), which dissolves completely once supersaturation diminishes. Surprisingly, the formation of the T-SEI promotes uniform morphological evolution during battery discharge through compact growth of the grains. These findings establish governing principles for T-SEI behavior, broadly applicable to interfaces with high solute production rates. This work highlights T-SEI's relevance to key electrochemical processes, including fast cathode charging and rapid anode discharging, offering new insights for advancing battery technologies.

**Figure 1. Classical Solid-Electrolyte Interphase (SEI) and Transient Solid-Electrolyte Interphase (T-SEI).** (A) Schematic diagram showing the formation of a classical non-transient SEI caused by electrolyte decomposition. Due to its heterogeneous nature, metal growth upon the next charge is non-uniform. (B) Schematic diagram showing the formation of a Transient SEI. Under fast discharging conditions, the dissolution of the metal results in a steep local concentration gradient near the electrode, leading to local supersaturation and salt crystallization. A flat, uniform surface is produced because of the self-limiting nature of T-SEI dominated dissolution. The T-SEI completely dissolves over a characteristic relaxation time $\tau_{R,SEI}$ when the local supersaturation is no longer present after the discharge is stopped.

**Figure 2. Understanding T-SEI formation dynamics using a Rotating Disk Electrode (RDE).** Current density-voltage ($J$-V) curves of electro-dissolution in (A) 20 mM, (B) 1 M, and (C) 3 M $ZnSO_4$ aqueous electrolytes. Time-dependent current measured in constant-potential electro-dissolution at different potentials with no rotation: (D) 1 M and (E) 3 M $ZnSO_4$ aqueous electrolytes. The working electrode is Zn metal foil. Scan rate is 100 mV/s for all measurements.

**Figure 3. Determination of the Critical Capacity $Q_c$ for T-SEI formation.** (A) Voltage profiles of electro-dissolution in 3 M $ZnSO_4$ at different current densities. (B) Critical capacity $Q_c$ dissolved before voltage spike plotted against reciprocal of current density for 3 M $ZnSO_4$ and 30 m $ZnCl_2$ electrolytes. (C) Optical images of the Zn metal working electrode at different times during 100 mA/cm$^2$ constant-current measurement (the green curve in panel a) in 3 M $ZnSO_4$ electrolyte.

**Figure 4. Probing T-SEI properties using Electrochemical Impedance Spectroscopy (EIS).** (A)-(B) EIS data represented on a Nyquist plot at different dissolution potentials. (C) Fitted values of charge transfer resistance as a function of potential and (D) fitted values of double layer

capacitance as a function of potential. The Zn metal anode is held at a series of potentials as detailed in the plots for the EIS measurements. Electrolyte: 3 M $ZnSO_4$.

**Figure 5. Surface characterization of the metal electrode after full relaxation of the T-SEI.** (A) Schematic diagram showing surface chemistry evolution associated with the T-SEI formation and relaxation. (B)-(C) FIB-SEM of interface in (B) Zn electrode discharged at high rate in 3.3 M $ZnSO_4$ and (C) pristine Zn. (D)-(F) EDX spectra of high-rate discharged (yellow), low-rate discharged (red), and pristine (blue) Zn electrodes showing oxygen (D), sulfur (E), and carbon (F) peaks. 1 mAh/cm$^2$ was discharged for both high-rate (5 V vs. $Zn^{2+}$/Zn) and low-rate (0.1 V vs. $Zn^{2+}$/Zn) samples. Electrolyte: 3 M $ZnSO_4$.

**Figure 6. Influence of T-SEI formation on electrodeposition morphology.** Scanning electron microscopy (SEM) images of (A)-(B) low rate discharged Zn electrodes at 0.1 V vs. $Zn^{2+}$/Zn and (C)-(D) high-rate discharged Zn electrodes at 5 V vs. $Zn^{2+}$/Zn after T-SEI formation and relaxation. Discharge capacity: 0.5 mAh/cm$^2$. SEM images of re-deposition morphology on (F-G) low-rate discharged and (H-I) high-rate discharged, T-SEI-formed/relaxed Zn electrodes. Re-deposition condition: 0.5 mAh/cm$^2$ at a constant potential of -1 V vs. $Zn^{2+}$/Zn. (I) Electron backscatter diffraction (EBSD) image of high-rate discharged Zn electrode. (J) EBSD image of the re-deposited Zn onto high-rate discharged, T-SEI-formed/relaxed Zn electrode. The flat, and coarse-grained morphology observed is fundamentally different from re-deposition on a low-rate discharged Zn electrode. All experiments were carried out in 3 M $ZnSO_4$.

**Main Text Begins:**

The invention and commercial use of metal anodes far preceded contemporary battery anodes based on ion-intercalation.[1] A metal anode utilizes the deposition/dissolution reaction—also called "plating/stripping" in a classical context—of a metallic element in the charging and discharging of the battery, respectively. The *in-situ* deposition/dissolution of a solid metallic phase inside a confined battery cell, unfortunately, turned out to be highly problematic, causing fast capacity fading, battery failure, and—in some cases even worse—fire hazards.[2, 3] In the past decade, there has been renewed interest in developing this type of metal anode.[4] Electrodes of this nature hold the potential to improve the energy density of batteries by several times or even up to one order of magnitude. A significant amount of research effort has been devoted to understanding the failure mechanisms of metal anodes, with an emphasis on critical interfaces/interphases developed during electrodeposition.

Upon contact between the metal anode and the electrolyte, a Solid-Electrolyte Interphase (**SEI**) is formed on the metal surface due to spontaneous redox reactions between the metal and the electrolyte. The formation of SEI on metal electrodes is present in aqueous, non-aqueous organic, and ionic-liquid electrolyte systems.[5-7] Leveraging state-of-the-art characterization tools, recent studies have shown the SEI exhibits highly complex and heterogeneous chemical, structural, and electrical properties.[8-12] This SEI heterogeneity dominates the electrodeposition morphology in subsequent cycles, creating undesirable porous metal deposits upon charging **(Fig. 1A)**. The porous metal deposits—on the one hand—are prone to mechanical break-off from the current collector resulting in isolated fragments known as dead or orphaned metals.[13-16]On the other hand, the high surface area of these porous structures further exacerbates electrolyte decomposition and SEI formation. It is now well established that SEI—a complex, thin passivating film consisting of inorganic and organic species—plays a pivotal role in the morphological evolution of electrochemically-deposited metals and, consequently, the rechargeability of the metal anodes.

The SEI formation and morphological evolution of metal anodes have been extensively studied in the context of *deposition*. In stark contrast, electrochemical *dissolution*—the reverse reaction that occurs during battery discharge—and its impact on the overall rechargeability of a metal anode remains poorly understood. A pioneering work based on focused-ion beam and scanning electron microscopy (FIB-SEM) shows that nanovoids form underneath the SEI, resulting in highly uneven dissolution morphology of a metal.[17] The uneven dissolution morphology dominates the subsequent electrodeposition upon battery recharge.[18, 19] Understanding the critical interfaces/interphases formed in dissolution and controlling the dissolution dynamics are needed to further advance metal anodes.

As a point of departure, we assess mass transport processes near a metal anode. Conventional wisdom holds that battery failure becomes inevitable when the cell is cycled at a current density $J$ approaching the diffusion limiting current density $J_{lim}$. In the deposition process, it is well known that the metal anode is susceptible to dendritic or porous growth at high current densities; This is driven by the steep concentration gradient formed due to the depletion of metal ions at the electrochemical interface at Sand's time $t_{Sand}$.[20] The mass transport of the dissolution process during battery discharge, however, is rarely evaluated, theoretically or experimentally. One might naively think that mass transport is an immaterial factor in dissolution; Dissolution releases solute metal cations into the liquid phase and may never result in a critical divergence, such as ion depletion at $t_{Sand}$ in deposition. This argument applies to a large body of classical electrochemical measurements in dilute electrolyte (~10 mM).[21, 22] It no longer holds true, however, for battery-relevant electrolyte systems where the salt concentration is greater than 1 M or much higher such as "water-in-salt" electrolytes. In analogy to the ion depletion limit in deposition, a steep local concentration gradient driven by dissolution in fast discharging could exceed the saturation limit of the salt and lead to precipitation of a passivating film on the electrode surface, cutting off the electrochemical reaction. As the battery research community turns to high concentration electrolytes,[23-26] thorough interrogation of the role played by mass transport in electrochemical dissolution becomes highly relevant and increasingly critical.

Using a rotating disc electrode (**RDE**) and *in-operando* visualization, we discover a Transient Solid Electrolyte Interphase (**T-SEI**) which forms on metal anodes due to salt supersaturation. The defining characteristic of T-SEI is that its presence is associated with a characteristic relaxation time $\tau_{R,SEI}$. The T-SEI can be fully relaxed—that is, to dissolve—over time. Our results further show that electrochemical deposition and dissolution surprisingly exhibit an "antisymmetric" relation; The mass transport limit—which destabilizes the dynamic interface in deposition (see **Fig. S1**)— stabilizes that interface in dissolution (see **Fig. 1B**). Finally, we found the formation of T-SEI has significant impacts on the deposition morphology achieved in the subsequent recharge. Flat, coarse-grained metal deposits are observed on metal anodes that underwent T-SEI formation, in stark contrast to isolated particulate metal deposits normally achieved in battery recharge.[27]

## Results and Discussion

### *Electrochemical formation of T-SEI in discharge*

We choose to study dissolution of Zn metal anodes in mildly acidic aqueous $ZnSO_4$ and $ZnCl_2$ electrolytes, respectively, as a model system. The reason is two-fold. Compared with Li, Zn has a moderate reduction potential and is less susceptible to spontaneous SEI formation due to electrolyte reduction. This allows us to definitively single out the mass transport in the liquid electrolyte and the resultant chemical dynamics in the dissolution process in question. In addition, Zn metal anodes in $ZnSO_4$ and $ZnCl_2$ have received growing research interest for their application in affordable, intrinsically safe electrochemical energy storage.[28] Unlike $Li^+$ ion, intercalation-type anodes for $Zn^{2+}$ ion are rare. Utilizing the deposition/dissolution reaction remains the most technologically-relevant solution for the anode in Zn-ion batteries.

We first use an RDE system to perform linear sweep voltammetry measurements of Zn dissolution in a series of electrolytes: 20 mM, 1 M, and 3 M aqueous $ZnSO_4$. In the dilute, 20 mM $ZnSO_4$, the dissolution of Zn shows purely classical ohmically-limited behavior (**Fig. 2A**). The effect of rotation makes no significant difference in the *J*-V signature. This suggests that mass transport is not the limiting factor in the dissolution of metal in a dilute concentration far away from the saturation limit. In the 1 M $ZnSO_4$ electrolyte, deviation from the ohmic linearity begins to be discernible at ~2.5 V followed by a significantly limited, plateauing behavior (**Fig. 2B**; See also **Note S1** on the magnitude of the measured voltages). This deviation is completely absent under a forced convection condition with 1000 rpm. Finally, in 3 M $ZnSO_4$, the mass transport effect is even more prominent (**Fig. 2C**). Beyond a critical potential of ~1.5 V, the current drops precipitously. The forced convection delays the deviation and lifts the plateau current density. Although the voltages used are in theory beyond the anodic stability of the electrolyte, it should be noted that most of the overpotential results from ohmic resistance in the electrolyte, as the data is not iR compensated (see **Note S1**). It should also be noted that no oxygen evolution reaction is observable throughout all measurements due to the extremely low catalytic activity of Zn with its fully filled 4s and 3d orbitals.[29] The apparent impact of electrolyte concentration and hydrodynamic field on the electrochemical signatures of dissolution strongly indicates the non-trivial role played by mass transport in dissolution dynamics.

The effects of mass transport can also be clearly seen in the constant-potential, chronoamperometry measurements of dissolution (**Fig. 2D** and **2E**). A large applied potential leads to a precipitous current drop, particularly in the 3 M $ZnSO_4$ solution, whereas a small applied potential leads to an initial increase in current due to surface roughening and removal of some native oxide species.[30-32] In fact, the rapid current decrease at large potentials in chronoamperometry is considered classical and is common in electrochemical reactions where the reacting agents are dissolved in the liquid electrolyte, such as electrodeposition. The decrease in current density over time is attributed to the consumption of reacting agents near the electrode and a growing diffusion boundary layer thickness, following $J(t) \propto \frac{1}{\sqrt{t}}$.[33] However, in dissolution, the reacting species is the metal itself that remains in excess throughout the measurements. The classical mass depletion theory, therefore, cannot explain the current drop over time. Intriguingly, the limiting behavior is more prominent in more concentrated electrolytes, on the contrary to what is expected for deposition. An analogous trend is observed for the dissolution of Zn in $ZnCl_2$ electrolytes (see **Fig. S2**). This observation is reminiscent of earlier findings in strongly alkaline Zn primary batteries during discharging.[34] The limiting behavior was attributed to the formation of a permanent ZnO interphase that prevents electrode self-discharge, due to supersaturation and local pH deviation from the strongly alkaline conditions.[35, 36] We hypothesize that the drop in current in the acidic $ZnSO_4$ and $ZnCl_2$ electrolytes is caused by the formation of a new type of passivating SEI that is produced as the local concentration exceeds the saturation limit. Further, the passivating SEI formed in acidic $ZnSO_4$ and $ZnCl_2$ electrolytes is anticipated to dissolve once the concentration gradient is relaxed.

To further understand the transport limited phenomena in dissolution, we performed constant-current chronopotentiometry measurements in tandem with *in-operando* visualization (**Fig. 3**). When the applied current density is greater than 60 mA/cm$^2$ in 3 M $ZnSO_4$, the potential measurements show a spike in the voltage of the working electrode after a critical time (**Fig. 3A**), which we will refer to as the T-SEI incubation time $t_{\mathrm{inc.}}$. This behavior of the potential is analogous to the classical Sand's time in electrodeposition; The voltage of the working electrode rapidly increases due to the concentration of the reacting species reaching zero near the electrode surface.[20] However, in this case the voltage spikes due to the formation of the passivating T-SEI discussed earlier. Analogous results have been seen in the electro-dissolution of copper in phosphoric acid baths, where the potential sharply increases due to local supersaturation and subsequent salt precipitation.[37] It should be particularly noted that the current density at which the T-SEI can be formed is similar to that used in common aqueous batteries: ~60 mA/cm$^2$ in 3 M $ZnSO_4$ or ~20 mA/cm$^2$ in 30 m $ZnCl_2$ (see **Fig. S3**), both of which are being explored as commercially-relevant electrolytes.[38-42] Therefore, when utilizing high concentration electrolytes that are close to the saturation limit, a rapid increase in the operation voltage during battery cycling cannot immediately be attributed to mass transport during deposition but instead could also be induced by the formation of the T-SEI. To confirm this, we assembled coin cells using a 3 M $ZnSO_4$ electrolyte and 1M $ZnSO_4$ electrolyte and ran a galvanostatic charge-discharge cycle at 20 mA/cm$^2$ and an areal capacity of 10 mAh/cm$^2$ (**Fig. S4**). A conspicuous voltage spike is observed in the coin cell using 3M $ZnSO_4$ electrolyte, but not in the coin cell using 1 M $ZnSO_4$ electrolyte. Comparing this observation in practical Zn||Zn symmetry cells, i.e., the absence of the sharp voltage increase in the less concentrated electrolyte, to results reported in **Fig. 2** and **Fig. 3**, we conclude that this electrochemical signature in coin cell cycling is a result of T-SEI formation. The results show that T-SEI is an overlooked yet critical factor in practical metal anodes in batteries.

To gain quantitative insights into the dynamics behind T-SEI formation, we plot the critical capacity $Q_c$ discharged upon voltage divergence versus the reciprocal of the current density $J^{-1}$ applied. A linear relation is found in the $Q_c$ versus $J^{-1}$ plot (**Fig. 3B**). Sand's analysis on electrodeposition, in surprising analogy, also predicts a linear relation between the critical capacity ($Q_{Sand}$, also known as the Sand's capacity, at which point the voltage diverges) and the reciprocal of current density.[20, 33, 43] In light of the analogy, we draw inspiration from the Sand equation to rationalize the observed linearity in dissolution. According to the Sand equation for a binary electrolyte, the incubation time before the concentration of the reactant reaches zero at the electrode surface can be described by $\frac{J\sqrt{t_{Sand}}}{c_i^{\infty}} = \frac{nF\sqrt{\pi D_i}}{2}$ (1) where $t_{Sand}$ represents the incubation time, $J$ represents the current density, $c_i^{\infty}$ is the bulk concentration of the electrolyte, and $D_i$ is the equivalent diffusion coefficient of the electrolyte salt, $n$ represents the number of electrons transferred, and $F$ is Faraday's constant (See **Note S2** for more information on the equivalent diffusion coefficient). The result is simply a consequence of Fick's second law and neglects the impact of migration and convection. In the case of electro-dissolution, equation (1) can be slightly modified to determine the incubation time it takes for the concentration to reach the saturation concentration at the electrode surface: $\frac{J\sqrt{t_{inc.}}}{c_i^{sat} - c_i^{\infty}} = \frac{nF\sqrt{\pi D_i}}{2}$ (2). After this incubation time, the T-SEI will precipitate and passivate the entire electrode surface. Solving for the discharge capacity, we find $Q_C = Jt_{inc} = \frac{\pi D_i\left((c_{sat} - c_i^{\infty})nF\right)^2}{4J}$ (3) or more simply, the discharge capacity is directly proportional to the reciprocal of the applied current density $J$. The observed linear relation between $Q_c$ and $J^{-1}$ in dissolution offers two important implications. First, similar to mass depletion in deposition, the voltage divergence is a manifestation of local steep concentration gradient. In dissolution, this steep gradient results in supersaturation, formation of T-SEI, and passivation of the electrode. Second, the nucleation time of the T-SEI is on a much smaller time scale than $t_{inc}$. The T-SEI forms almost instantaneously upon supersaturation.

*In-operando* optical images of the metal electrode during chronopotentiometry (**Fig. 3C**) reveal the formation process of the T-SEI. It propagates from the edges of the electrode and then completely covers the surface of the foil at which point the voltage spikes. See **Fig. S5** for more details on the setup used for taking the optical images. A few earlier studies in the context of electrochemical polishing and machining examined the dissolution of iron in highly concentrated $FeCl_2$ electrolyte. It was hypothesized that the salt film consisted of two parts: a highly compact film on the order of tens of nanometers thick, and a highly porous film which is on the order of microns thick [44]. In light of the visibility, the propagating interphase seen in **Fig. 3C** should belong to the latter. The propagation of the T-SEI from the edges can be rationalized by considering the current density, and therefore the salt concentration, is concentrated at the edges of the foil from Laplace's equation $\nabla^2 \emptyset = 0$ (4).[45] This set of *in-operando* visualization results conclusively establishes the direct correlation between the formation of T-SEI and the observed, transport-limited electrochemical signatures in **Fig. 2** and **3**.

To quantitatively probe the properties of the T-SEI and further understand the observed electrochemical phenomena, we use potentiostatic electrochemical impedance spectroscopy (EIS). **Fig. 4A** and **4B** show the difference between the EIS signatures obtained at low dissolution overpotential and high dissolution overpotential on a Nyquist plot. Equivalent circuit modeling can give information on the ohmic resistance $R_u$, charge transfer resistance $R_{ct}$, and double layer capacitance $C_{dl}$. We found that increasing the voltage beyond a critical voltage during dissolution results in a dramatic increase in $R_{ct}$ (**Fig. 4C**). Thus, the electrochemical signatures seen in **Fig. 2** and **3** can be attributed to the dramatic increase in kinetic impedance upon T-SEI formation. Classically, $R_{ct}$ is expected to decrease with increased overpotential due to the exponential form of the Butler-Volmer equation;[46] This trend is observed in the low overpotential range (**Fig. 4B**). The analysis on $R_{ct}$ suggests that the T-SEI gives rise to a different kinetic mechanism for dissolution than in the classical solid-liquid interface that exists at low dissolution overpotentials. The result is in stark contrast to earlier work on electropolishing, which hypothesized the film formed upon local supersaturation was of a completely resistive nature.[37]

In addition, $C_{dl}$ of the interface dramatically decreases to several orders of magnitude lower than that expected for a normal aqueous interface (**Fig. 4D**).[47] The result can be understood by examining the physical picture of the capacitive process. As the thickness of a capacitor increases, its capacitance decreases proportionally as the electric field needs to transverse a longer distance. Therefore, the T-SEI can be thought of as its own capacitor which is much thicker than the electrical double layer that occurs in a conventional aqueous electrolyte. The EIS analyses are consistent with earlier reported on iron and copper for electro-polishing.[44, 48] The significant change in $C_{dl}$, as well as the visible interphase seen in **Fig. 3** definitively confirms the formation of a T-SEI that changes the electrochemical signatures.

We also notice that the ohmic resistance increases by ~10% at higher potentials (**Supplementary Table S1**). We attribute this to the low ionic conductivity of the T-SEI, which consists mainly of $ZnSO_4$. However, because the layer is thin, likely on the order of tens of nanometers,[44] the ohmic resistance only increases slightly, given $R = \frac{l}{A}\rho$, where $R$ is ohmic resistance, $l$ is transport length, $A$ is cross-sectional area, and ρ is resistivity.

***Quantifying the transient nature of T-SEI***

In theory, this interphase starts to dissolve once the concentration gradient near the electrode is relaxed and approaches the bulk concentration. To characterize the transient nature of the T-SEI, we repeat constant-current chronopotentiometry measurements between different rest intervals $t_{rest}$ in 3 M $ZnSO_4$ and determine how the length of the rest interval $t_{rest}$ after the first measurement affects the subsequent measurement (**Fig. S6**). If the interphase is "permanent", *i.e.*, $\tau_{R,SEI} \rightarrow \infty$, the subsequent measurement should immediately report a divergent, large voltage ($t_{inc,2} = 0$), regardless of the magnitude of $t_{rest}$. On the contrary, if the interphase is transient, when $t_{rest} \gg \tau_{R,SEI}$, the T-SEI has fully dissolved and the subsequent measurement will have a nonzero $t_{inc,2}$. See **Table 2** for a definition of these terms. The results show that, in 3 M $ZnSO_4$, $\tau_{R,SEI}$ is at most 300 milliseconds, given the nonzero $t_{inc,2}$ (**Fig. S6C**). We define a second relaxation time, $\tau_{R,conc.}$, which is the relaxation time for the concentration gradient formed in the electrolyte after dissolution. This determines the dependence of $t_{inc,2}$ on $t_{rest}$, i.e., a smaller $t_{rest}$ will decrease the value of $t_{inc,2}$ because the concentration at the electrode surface is still significantly above the bulk electrolyte concentration **(Fig. S7)**. The 30 m $ZnCl_2$ electrolyte has a $\tau_{R,conc.}$ on the order of $10^2$ seconds, while the 3 M $ZnSO_4$ has a $\tau_{R,conc.}$ on the order of $10^1$ seconds **(Fig. S7B and S7C)**. Comparing both time scales ($\tau_{R,SEI}$ and $\tau_{R,conc.}$) to a normal battery operation cycle (*e.g.*, 1 C, 1 hour for charge and discharge, respectively), the phenomenon of T-SEI formation and dissolution—despite the dominant role it plays at the dynamic electrochemical interface—is of a highly transient nature (**Fig. 5A**). As a result, the T-SEI is not detectable in most *post-mortem* materials characterization techniques. The complementary methodology employed here, consisting of *in-operando* EIS / visualization and other electroanalytical measurements, unambiguously reveals the presence and properties of T-SEI.

We use state-of-the-art focused-ion beam (FIB) to create a "trench" on the metal electrode and expose the cross-section of the surface. The cross-section is then studied by scanning electron microscopy (SEM) and energy dispersive X-ray spectroscopy (EDS). Comparing the surface of a pristine and a T-SEI formed electrode, no permanent interphase is detectable after T-SEI formation; The SEM and EDS only show the platinum protective layer deposited during the FIB and the bulk Zn (**Fig. 5B** and **5C**; **Fig. S8**). The complete absence of T-SEI in this *post-mortem* FIB-SEM/EDS analysis further corroborates its unusual transient nature. *In-operando* techniques with high temporal resolution (below $\tau_{R,SEI}$) are required to detect the T-SEI structurally or chemically.[49]

The surface chemistry of the electrode is, however, strongly influenced by the discharge conditions (**Fig. 5D~5F**). The pristine, as-received commercial metal electrode shows a significantly higher C content, indicative of adventitious carbon residue covering the surface. The metal electrode discharged at a low rate, well below the T-SEI formation regime shows a significantly reduced C content but increased S and O contents; This suggests the C residue layer is replaced by a S/O-enriched layer, likely a zinc hydroxysulfate or derivatives, in low-rate discharging.[50] It also creates the so-called corrosion "pits" formed unevenly on the electrode surface (**Fig. 6A** and **6B**). The pits grow significantly in size as a higher capacity is discharged, and corrosion by-products can be seen formed inside the pits (**Fig. S9**). Interestingly, a T-SEI formed metal electrode shows reduced surface contents of C, S, and O, compared with the other two samples (**Fig. 5D~5F**). This means that the T-SEI formation process chemically "cleans" the metal surface by removing the C residue layer while avoiding excessive formation of hydroxyl sulfate (**Fig. 5A**).

***Unusual morphological evolution promoted by T- SEI***

Next, we study the impact of T-SEI formation on the electro-deposition/dissolution processes. The morphological evolution in electro-deposition/dissolution is highly sensitive to the dynamics at the interface. In electro-deposition, morphology obtained in a transport-limited, ion-depleted regime exhibits ramified, dendritic patterns and non-uniformity across the electrode. To our surprise, the electrode surface morphology formed under T-SEI regulation displays high homogeneity across the electrode (**Fig. 6C**) with fine sub-micron scale features (**Fig. 6D**). Conventional wisdom holds that the stabilizing effect of high currents results from more uniform pit nucleation due to greater overpotentials.[51] However, this hypothesis does not explain the critical transition we observed when the applied current density exceeds the transport-limiting current density. Electron backscatter diffraction (EBSD) of the sample unveils the grain size and shape of the metal electrode (**Fig. 6I**). 3-D profilometry across a 1.4 by 1.4 mm area shows that the roughness of the T-SEI formed electrode is lower than the low-rate dissolved electrode (**Fig. S10**). Analogous morphological observations are seen in 1M $ZnSO_4$ electrolyte and 30 m $ZnCl_2$ electrolyte (**Fig. S11**, **Fig. S12**). The even morphology developed under transport-limited dissolution and the rough morphology under transport-limited deposition constitute an "anti-symmetric" relation; That is, the mass transport limit results in self-reinforced, interfacial instability in deposition, but self-limited, interfacial stability in dissolution (**Fig. S13**).

Finally, we investigate the influence of T-SEI on the re-deposition process in the subsequent recharge. We deposit metal back onto the electrodes after dissolution and then use SEM-EBSD to study the morphology and crystallography of the new deposits. Deposits obtained on the low-rate discharged electrode show three-dimensional particulate nuclei with a diameter of 2-3 μm (**Fig. 6E** and **6F**). Electrodeposition onto pristine, as-received commercial electrode also results in an independent nucleation mechanism (**Fig. S14B**). In addition, low-current redeposition also results in an independent nucleation mechanism, so the heterogeneous morphology of the deposits cannot be attributed to mass transfer limitations (**Fig. S15A**). Conversely, the deposits on the electrode discharged at a high-rate are homogeneous and show characteristic crystallographic terrace features, providing important implications on the grain structure.[52, 53] Based on the SEM metallography, the grain size of the re-deposited metal is around 15~20 μm (**Fig. 6G** and **6H**; See also **Fig. S16** for close-ups). We further performed EBSD on the re-deposited surface (**Fig. 6J**), which shows a comparable grain size with the electrode before re-deposition (**Fig. 6I**). More evidence showing the role of T-SEI in inducing compact growth of the grains can be seen in **Fig. S17 – Fig. S20**.

These analyses suggest that the growth that happens on a T-SEI formed electrode is completely different from that happens on a normal electrode. Based on the observations, we hypothesize that metal re-deposition on a T-SEI formed electrode occurs primarily via compact growth of the original coarse grains in the electrode, whereas re-deposition on a normal electrode surface involves independent nucleation that is not controlled by the underlying metal electrode. This can be rationalized by the "cleaning" effect of the T-SEI in discharging and the suppression of excessive formation of zinc hydroxysulfates (**Fig. S21**), as discussed in **Fig. 5D~F**. Similar compact growth of existing grains is also reported in other systems when the surface passivation layer is removed.[54] In prior literature reports, *ex-situ* chemical treatment, polishing, or electrolyte additives are required to achieve such deposition.[55-57] The present work demonstrates that compact growth of grains, as opposed to independent nucleation, can also be attained by controlling the previously overlooked T-SEI. The uncommon growth mode is considered an approach towards flat and stable electro-deposition/dissolution morphology and high electrode rechargeability.

Motivated by this finding, we explore the potential of T-SEI as a new tuning knob for regulating the long-term morphological evolution in metal anodes over practical cycling. We first found that the initial compact growth mechanism observed after T-SEI formation is not self-sustainable over cycling. Large capacities of deposition reveal the transition into an independent nucleation mechanism (**Fig. S22A**), and large capacities of stripping at low rates after T-SEI formation show a heterogeneous morphology (**Fig. S22B**). This may be attributed to defects and impurities buildup.[58-62] In light of the gradual degradation over cycling after T-SEI formation, we hypothesize that T-SEI needs to be enforced periodically in the electrode cycling protocol to sustain the flat morphology. To test this, we constructed symmetric Zn cells. The control cell is cycled at 20 $mA/cm^2$ for 11 cycles, whereas the proof-of-concept cell is cycled using the same current density but adopts a fast-discharge step every five cycles to intentionally induce T-SEI formation. The resultant morphology of the electrodes is analyzed using a 3D surface laser profilometer. Using the T-SEI formation protocol results in a 42% reduction in surface roughness of the Zn foil compared to the control cell cycled under normal constant current conditions (**Fig. S23**). In addition, monitoring the potential at the beginning of each deposition step shows the overpotential for deposition is larger immediately after each T-SEI formation step (**Fig. S24**). The increased overpotential indicates reduction of surface roughness after T-SEI formation. The flat electrode surface after T-SEI formation lowers the electrochemically active area on the electrode surface; A larger overpotential is needed to achieve the same current density. The results obtained with this simplistic, proof-of-concept cycling experiment showcase the new design space for metal anodes opened by understanding and controlling T-SEI.

## Conclusion

The investigation into the formation, structure, and properties of SEI in electrodes has been a pivotal topic in contemporary battery research. The observation of T-SEI adds a new perspective to the existing paradigm of SEI. The results presented conclusively show that the T-SEI determines not only the electrochemical responses of the metal anode upon discharge, but also the long-term morphological evolution. We emphasize the extraordinary relevance of T-SEI for emerging electrolytes with relatively high salt concentrations, where the T-SEI forms at moderate current densities such as 20 $mA/cm^2$ or even smaller in low-diffusivity systems. In addition to metal anodes, any electrode processes that produce cations (*e.g.*, fast charging of an intercalation-type cathode, fast discharging of an intercalation type anode), in theory, are susceptible to T-SEI formation. We note, however, these processes are also limited by solid-state diffusion of ions—for example—through the permanent SEI on an alkali-metal surface or the bulk of an active particle in an intercalation electrode. We conclude that understanding and controlling T-SEI constitute a fresh line of inquiry for electrochemical engineering in next-generation battery systems with exceptional rate capabilities.

**Materials & Methods**

Materials and preparation of electrolytes

Zn foil (99.9%, 0.1 mm thickness) was purchased from MSE supplies. $ZnSO_4 \cdot 7H_2O$ (≥ 99%, ReagentPlus) and $ZnCl_2$ (98-100.5%, puriss. grade) were purchased from Sigma-Aldrich. All chemicals and metal foils were used as received. All electrolytes were prepared within vials under the ambient air by weighting certain amounts of salts to be dissolved into deionized water.

Electrochemical Measurements:

Square Zn foils were attached to a Rotating Disc Electrode with a Glassy Carbon interchangeable tip (Pine Research Inc.) using conductive silver paste (Delta-Technologies). The setup was heated for at least 30 minutes at 90 degrees Celsius for the silver paste to become active and the ohmic resistance between the base of the disc electrode and the foil was tested using a multimeter (Fluke Corporation) to ensure Zn foil was electrically connected to the glassy carbon. Then, the edges of the square foil were passivated using Kapton tape to minimize edge effects, and the final area for calculating the current density was measured with a ruler. All electrochemical measurements were done using a three-electrode setup with a Zn reference and counter electrode in a glass RDE cell (Pine Research) and conducted using a Biologic SP-300 instrument. Linear Sweep Voltammetry measurements were conducted at a scan rate of 100 mV/s. During subsequent electrochemical experiments in the same electrode, the electrode was allowed to rest until the open circuit potential fully stabilized at a normal value. PTFE plugs and parafilm were used to minimize the impact of evaporation during subsequent electrochemical measurements. During Electrochemical Impedance Spectroscopy, chronoamperometry was done at the specified bias voltage until the current reaches a quasi-steady state, at which time the EIS measurements started. EIS measurements were conducted at frequencies of 7 MHz to 1 Hz with an AC amplitude of 10 mV.

Materials Characterization:

EDX, SEM, FIB-SEM, EBSD, and 3-D Profilometry were used for characterizing the samples ex-situ. Before using these measurement tools, samples were washed with DI water and dried with a Kimwipe to get rid of any residual electrolyte. SEM images were taken on an Apreo 2C Lovac (Thermo Scientific) using an Everhart-Thornley detector (ETD). For obtaining EDX spectra, an accelerating voltage of 30 kV was used and data was collected for 3 minutes. The experiment was repeated four times on each sample to ensure repeatability and the total counts from the resulting Zn peak were compared across samples (**Fig. S25**) to ensure it was reasonable to compare counts from other elemental peaks.

For obtaining the FIB-SEM image, a Scios 2 HiVac Dual Beam was used. Coincidence between the ion beam and the electron beam was set by selecting a eucentric height. The stage was tilted to 52 degrees, then the Ga-ion-beam was used to deposit a Platinum protective layer. After focusing, a 10 µm cross section was milled using the ion-beam and then imaged.

For obtaining the EBSD images, the sample was inserted on a 45-degree line and the stage was tilted 25 degrees so the angle between the EBSD detector and the sample was 70 degrees. A 20 kV accelerating voltage was applied with a current 26 nA. The sample was focused using SEM before conducting EBSD. 20 background images were collected before calibrating the pattern center. Then, an exposure time of 20 ms and a gain of 20 was used to obtain the final grain maps. For obtaining 3-D Profilometry, a Keyence VK-3000 was used. The data was then analyzed in Multi-File Analysis Application Software. The image was then processed using the Surface Shape

Correction to account for any shape variations resulting from the foil not being perfectly flat. Finally, the $S_{dr}$ was calculating using the Multi-File Analysis Application Software to quantitatively evaluate the surface roughness. It is evaluated using $S_{dr} = \frac{1}{A}\iint\left(\sqrt[2]{1+\left(\frac{\partial z(x,y)}{\partial x}\right)^2+\left(\frac{\partial z(x,y)}{\partial y}\right)^2}-1\right)dxdy$ (4), where the integral over the entire surface area.

For operando optical imaging of the electro-dissolution process, a Skybasic wireless digital microscope was used.

<u>Cell Construction:</u>

To construct CR 2032-type Zn-Zn symmetric coin cells, stainless steel casing was used with two layers of polypropylene separator and 200 μL of electrolyte solution. Square foils were cut out with an area of 0.5 $cm^2$. The cells were cycled at a current density of 20 $mA/cm^2$ for an areal capacity of 10 $mAh/cm^2$. A five second rest was applied between each galvanostatic hold. A hydraulic press was then used to crimp the coin cell. Coin cells were cycled on a Neware battery cycler.

To construct the symmetric glass cell, a Zn foil with only 1 $cm^2$ of exposed surface area was immersed in a glass vial along with a Zn counter electrode with an exposed surface area of approximately 8 $cm^2$. 3 M $ZnSO_4$ electrolyte was then poured in the glass vial and it was covered with a slip of parafilm to minimize the effect of evaporation. To induce T-SEI formation during cell cycling, a special step protocol during the stripping process of every five cycles: 1.25 $mAh/cm^2$ was stripped off the anode galvanostatically at a current density of 20 $mA/cm^2$, and 0.75 $mAh/cm^2$ was stripped off the anode by holding the anode potentiostatically at 5 V versus the counter electrode. The cell was then rested for 3 minutes for the open circuit potential to stabilize. This special protocol during stripping of the working electrode was done every five cycles. All other steps were galvanostatic plating and stripping at a current density of 20 mA/cm$^2$ for an areal capacity of 2 mAh/cm$^2$, with 10 seconds of rest in between. The control cell used this normal protocol for the entire duration of cycling.

## Acknowledgements

The authors acknowledge financial support from Department of Defense National Defense Science and Engineering Graduate Fellowship (DoD NDSEG) and the University of Texas Science and Technology Acquisition and Retention (STARs) Fund. The authors thank Raluca Gearba for assisting in the FIB-SEM measurements. The authors also thank Rustam Gandhi, Wen-Yang Jao, and Evalyn Wilber for many useful discussions. **Author Contributions:** K.Z. directed the research. S.F. and K.Z. conceived and designed this work. S.F. and K.Z. wrote the paper. S.F. performed the electrochemical measurements and materials characterization. S.F. and K.Z. analyzed the data.

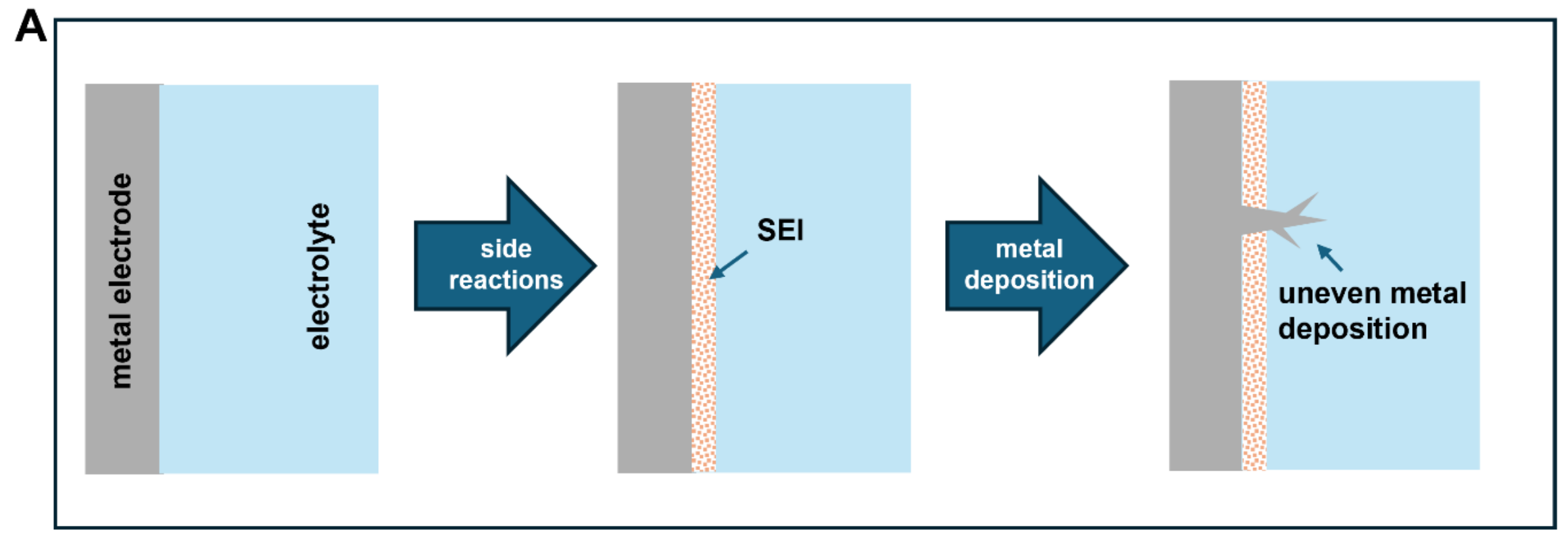

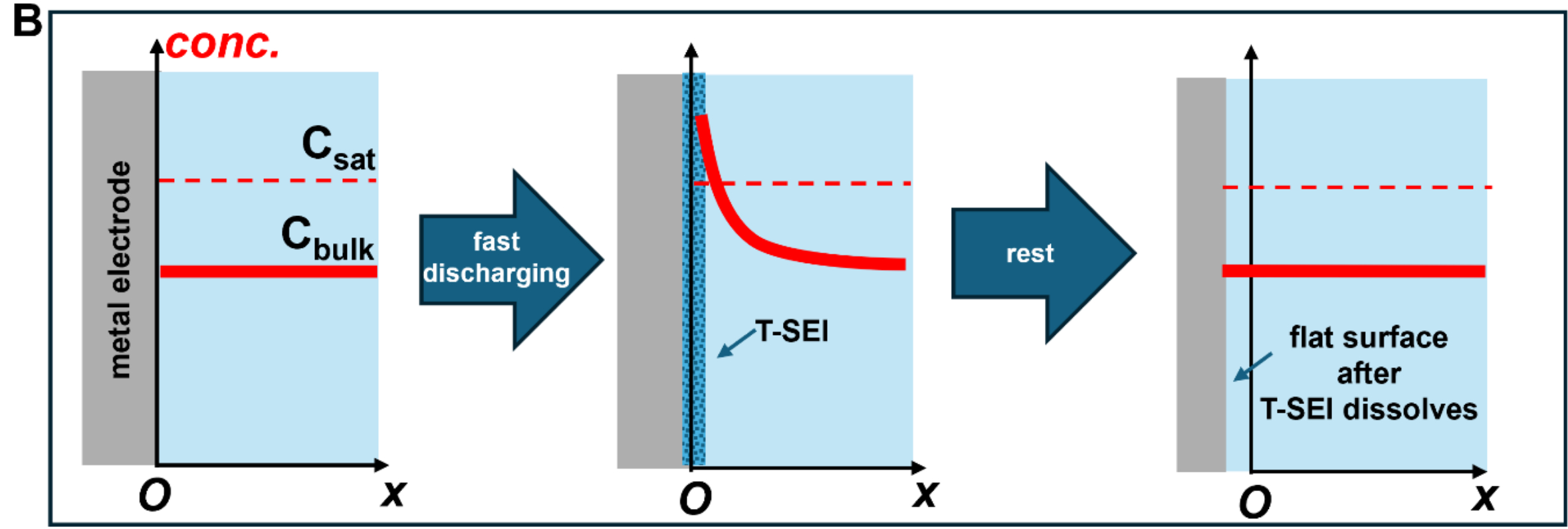

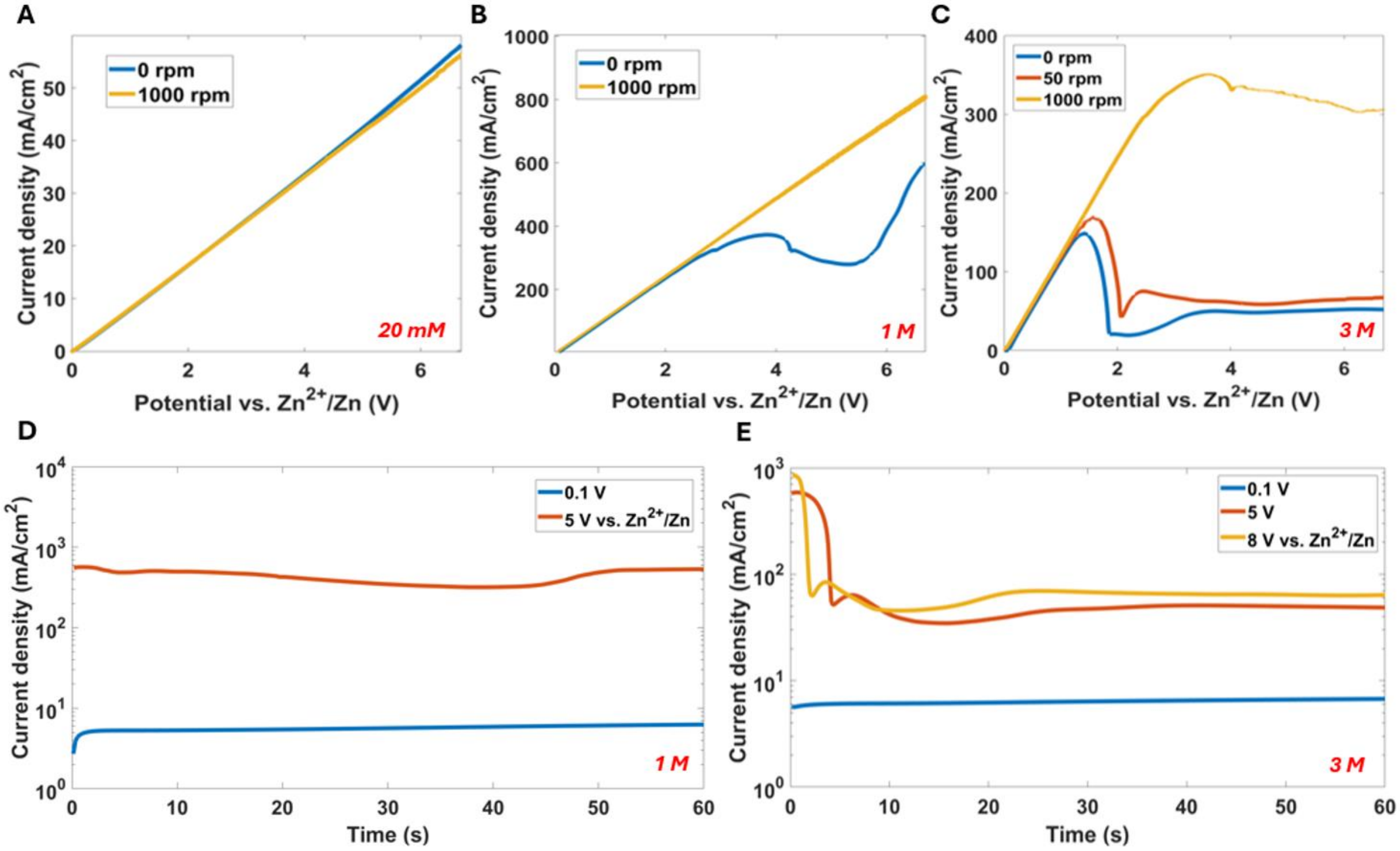

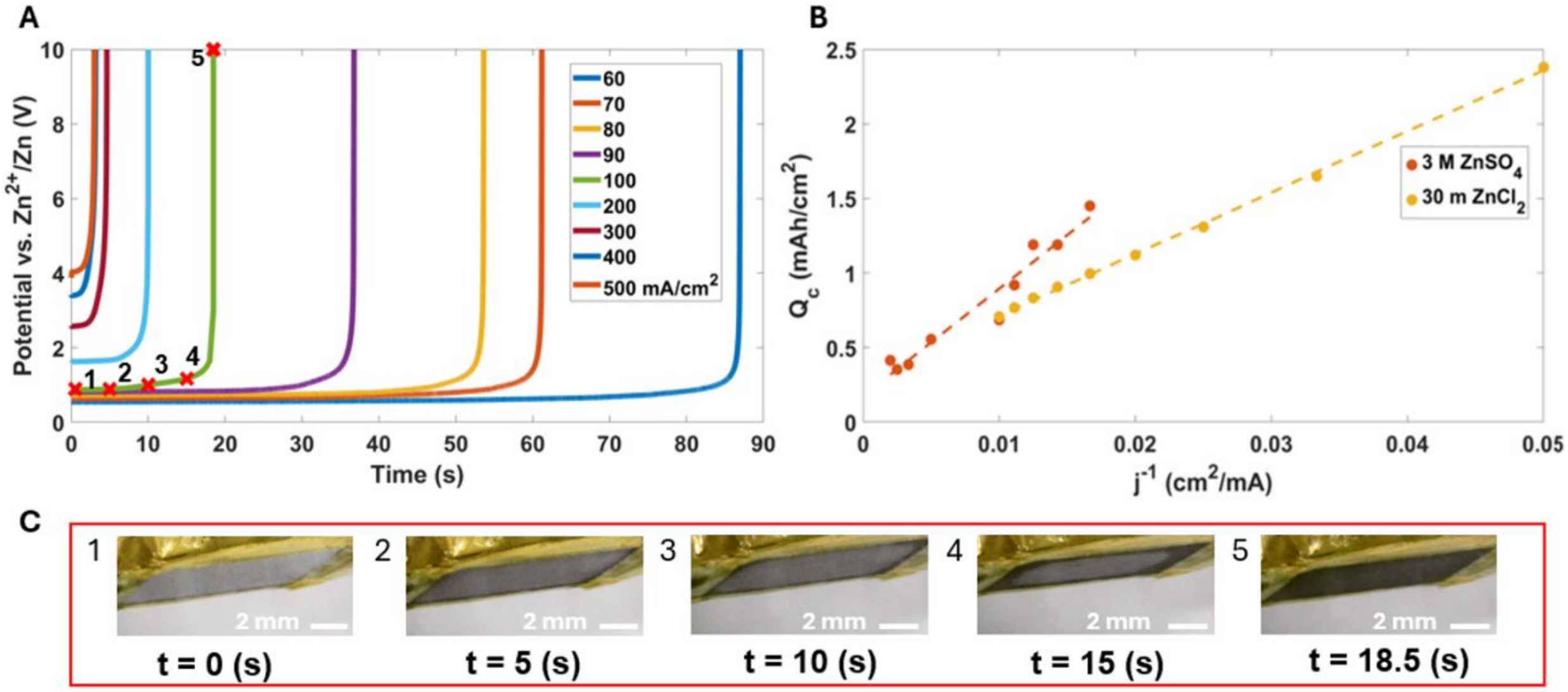

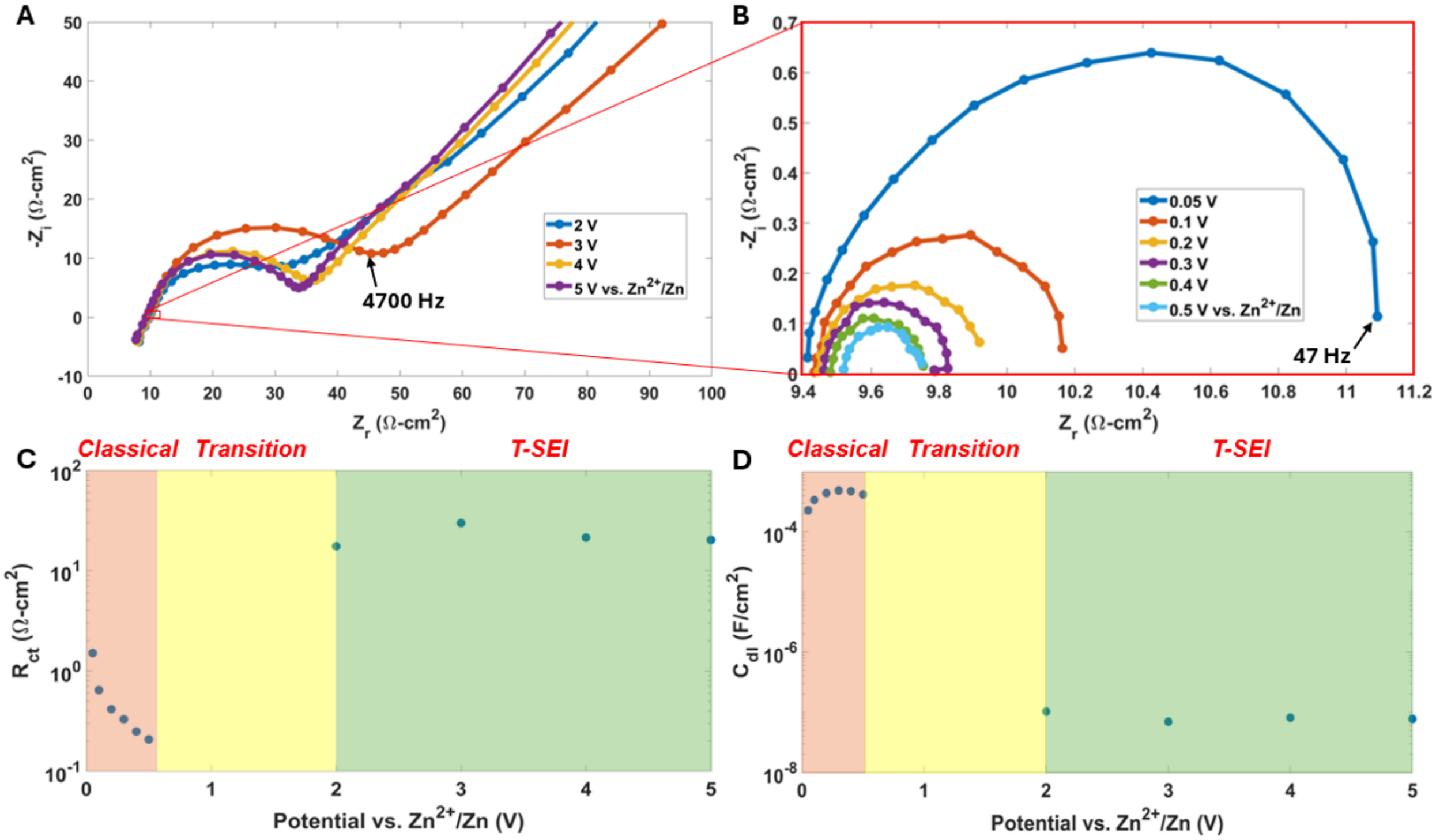

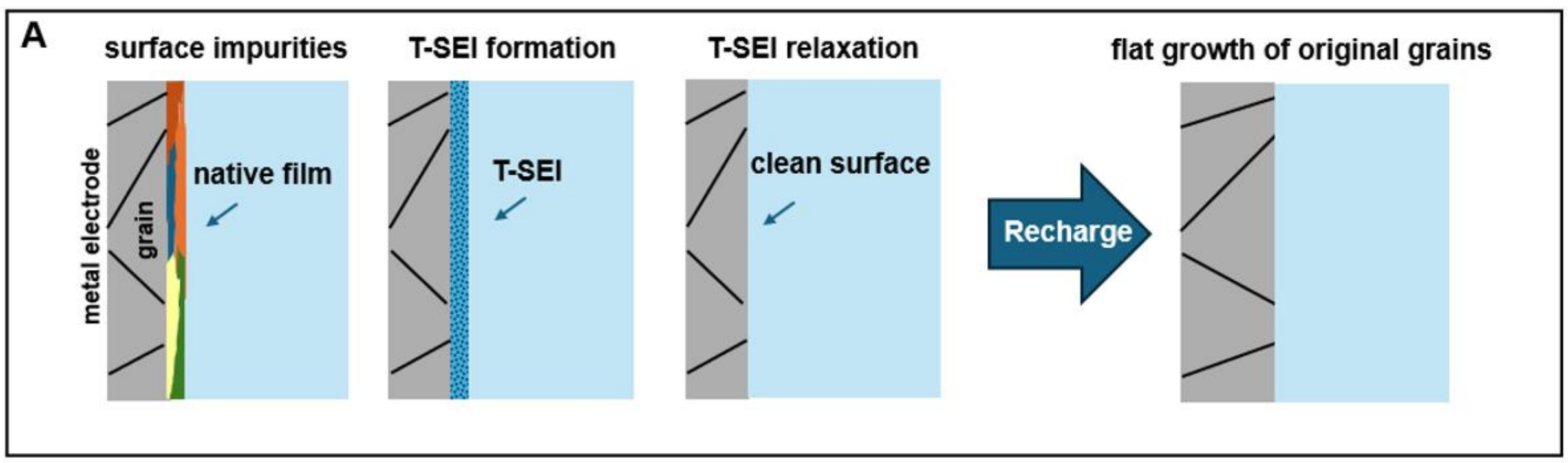

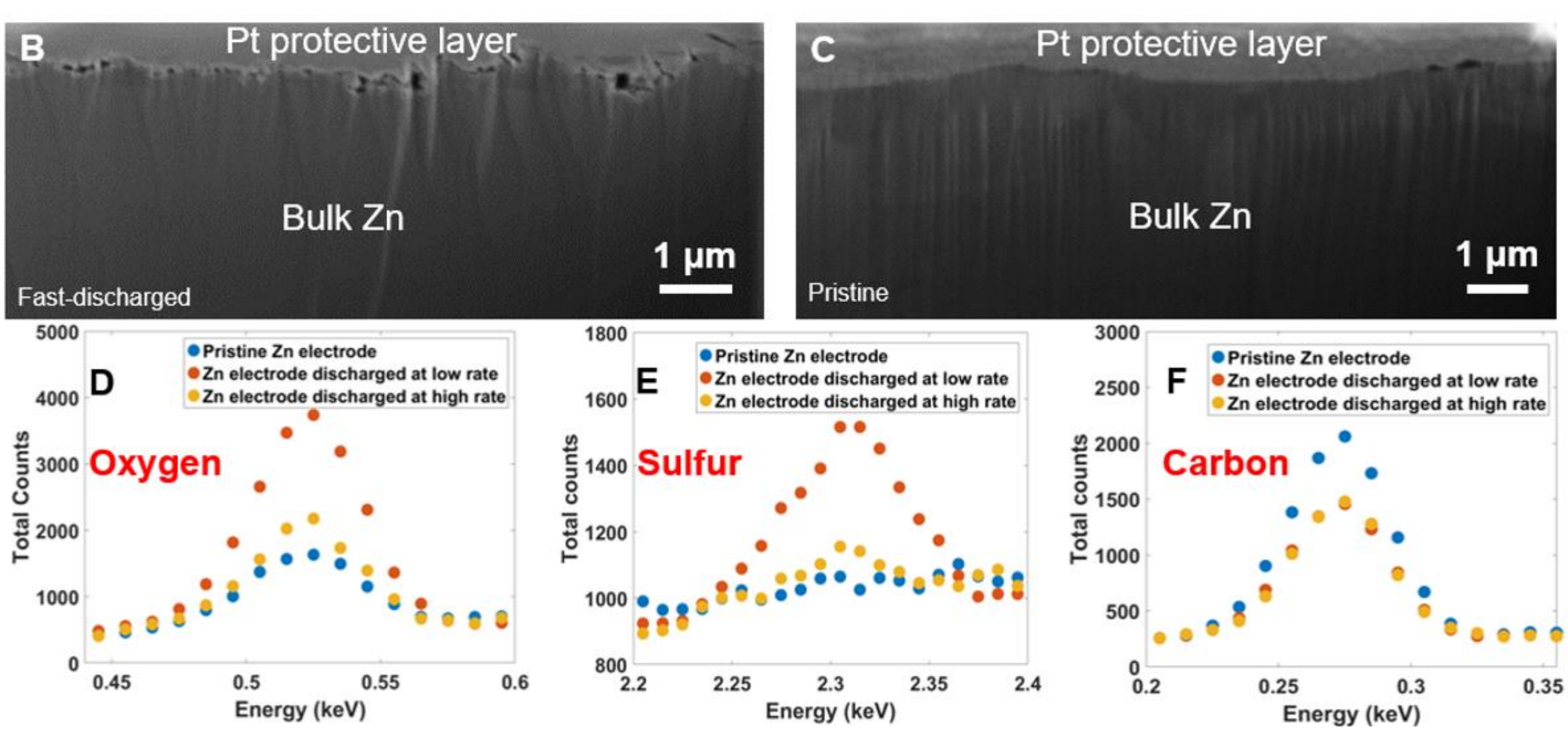

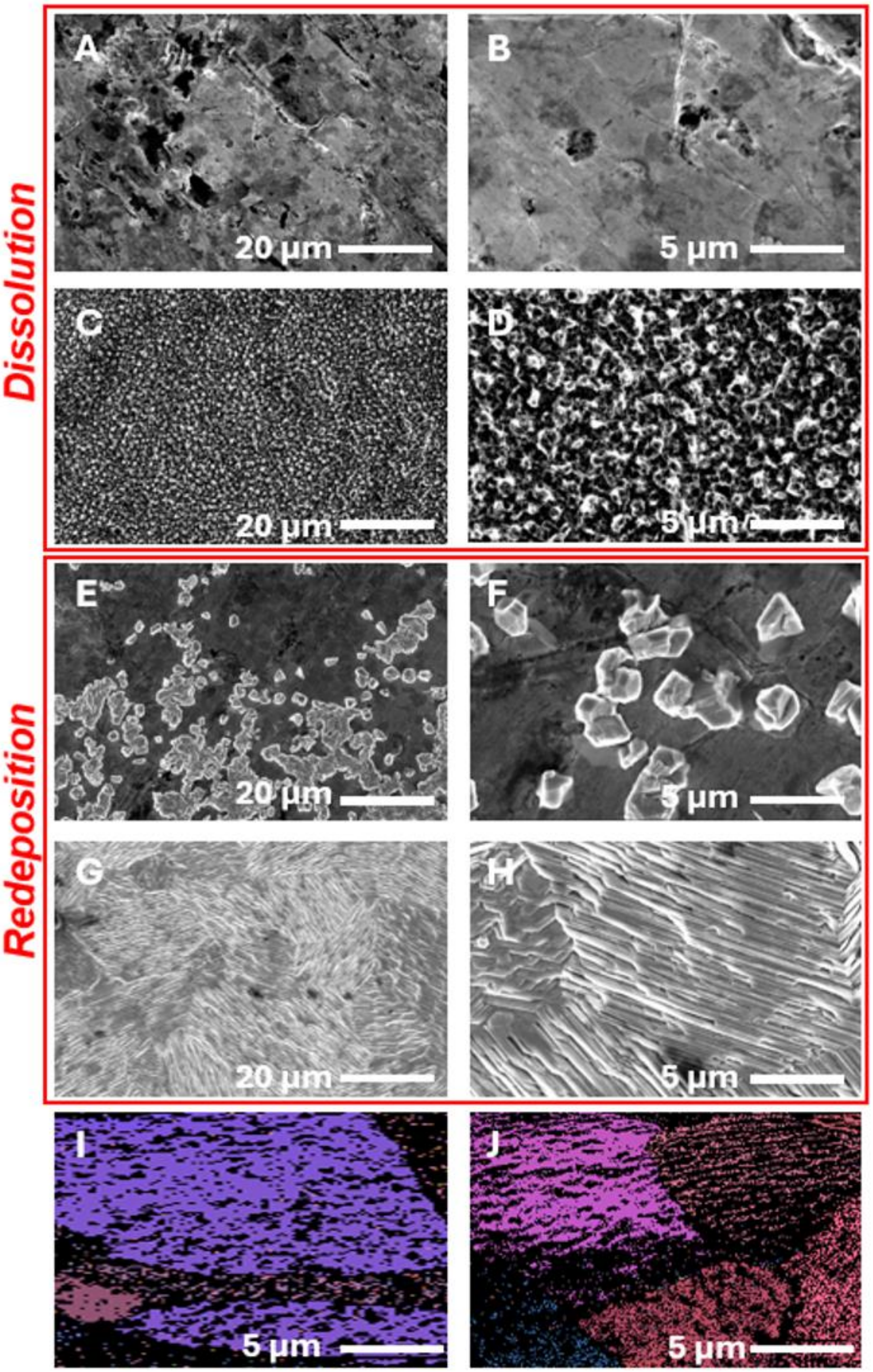